\documentclass[sigconf]{acmart}

\usepackage{amsmath}

\usepackage{amssymb}
\usepackage{booktabs}
\usepackage{graphicx}
\usepackage{array}

\graphicspath{{Materials/}}

\renewcommand\footnotetextcopyrightpermission[1]{}
\acmConference[ACM SIGSPATIAL '26]{34th ACM International Conference on Advances in Geographic Information Systems}{November 2026}{Riverside, California, USA}
\acmYear{2026}
\copyrightyear{2026}

\begin{document}

\title{GIScholarBench: Benchmarking LLM Overconfidence in GIS Research [Experiment]}

\author{Zongrong Li}
\authornote{Both authors contributed equally to this research.}
\affiliation{%
  \position{zongrong@tamu.edu}
  \department{Department of Geography}
  \institution{Texas A\&M University}
  \city{College Station}
  \state{TX}
  \country{USA}}

\author{Mingzheng Yang}
\authornotemark[1]
\email{ymz2020@tamu.edu}
\affiliation{%
  \department{Department of Geography}
  \institution{Texas A\&M University}
  \city{College Station}
  \state{TX}
  \country{USA}}

\author{Lei Zou}
\authornote{Corresponding author.}
\email{lzou@tamu.edu}
\affiliation{%
  \department{Department of Geography}
  \institution{Texas A\&M University}
  \city{College Station}
  \state{TX}
  \country{USA}}

\author{Hongxu Ma}
\email{hxma@google.com}
\affiliation{%
  \institution{Google}
  \city{Mountain View}
  \state{CA}
  \country{USA}}

\author{Hao Tian}
\email{haotian@tamu.edu}
\affiliation{%
  \department{Department of Geography}
  \institution{Texas A\&M University}
  \city{College Station}
  \state{TX}
  \country{USA}}

\author{Siqi Zhou}
\email{siqizhou1107@tamu.edu}
\affiliation{%
  \department{Department of Geography}
  \institution{Texas A\&M University}
  \city{College Station}
  \state{TX}
  \country{USA}}

\author{Wenjing Gong}
\email{wenjinggong@tamu.edu}
\affiliation{%
  \department{Department of Landscape Architecture and Urban Planning}
  \institution{Texas A\&M University}
  \city{College Station}
  \state{TX}
  \country{USA}}

\author{Kaili Zhang}
\email{kelly_zhang@tamu.edu}
\affiliation{%
  \department{Department of Landscape Architecture and Urban Planning}
  \institution{Texas A\&M University}
  \city{College Station}
  \state{TX}
  \country{USA}}

\author{Bingqian Chen}
\email{bingqc@tamu.edu}
\affiliation{%
  \department{Department of Geography}
  \institution{Texas A\&M University}
  \city{College Station}
  \state{TX}
  \country{USA}}

\author{Mitch Zhang}
\email{mzhang22@tamu.edu}
\affiliation{%
  \department{Department of Geography}
  \institution{Texas A\&M University}
  \city{College Station}
  \state{TX}
  \country{USA}}

\author{Yifan Yang}
\email{yyang295@tamu.edu}
\affiliation{%
  \department{Department of Geography}
  \institution{Texas A\&M University}
  \city{College Station}
  \state{TX}
  \country{USA}}

\begin{abstract}
Large language models (LLMs) are increasingly embedded in academic research workflows, yet the high factual precision required in scholarly tasks makes them especially vulnerable to overconfidence. We use this term in a behavioral sense---the tendency to produce confident, assertive, and well-formatted outputs even when the underlying knowledge is incomplete or unverifiable---rather than in the calibration sense of a measured mismatch between a model's self-reported confidence and its accuracy. To systematically evaluate this problem, we present GIScholarBench, a benchmark constructed from 10,865 papers collected from 25 core GIScience journals published between 2020 and 2025. The benchmark includes three tasks of increasing cognitive complexity: metadata retrieval, literature linking, and research direction generation. We evaluate Claude Sonnet 4.5, Gemini 3, and ChatGPT 5.3 through their native web interfaces under real-world user-facing conditions. The results reveal consistent overconfidence across all three tasks. In metadata retrieval, ChatGPT 5.3 achieves the highest accuracy (0.61--0.87), yet all models continue generating definitive titles and DOIs even when predictions are incorrect. In literature linking, Claude Sonnet 4.5 recovers the most references (Avg Hits = 10.3, P@20 = 0.53), but all models exhibit a large divergence between Hit@1 (0.70--0.97) and Hit@10 (0.08--0.61), indicating that citation lists are extended beyond reliable retrieval capacity. In research direction generation, AI-generated directions achieve substantially lower Topic Coverage (0.09--0.12) than real future-citing papers (0.22), with Novel Miss Rates of 0.88--0.91 and lower semantic diversity. These findings suggest that LLM overconfidence is task-invariant but form-varying, appearing as factual overgeneration in retrieval, unreliable citation expansion in literature linking, and overconfidence in output completeness in research ideation.
\end{abstract}

\ccsdesc[500]{Computing methodologies~Natural language processing}
\ccsdesc[300]{Computing methodologies~Machine learning}
\ccsdesc[300]{Information systems~Information retrieval}
\ccsdesc[300]{Information systems~Geographic information systems}
\ccsdesc[100]{General and reference~Evaluation}

\keywords{Large language models; Overconfidence; GIScience; Benchmark evaluation; Citation hallucination; Research direction generation}

\renewcommand{\shortauthors}{Li et al.}

\maketitle

\section{Introduction}

Large language models (LLMs) have become increasingly embedded in scientific research workflows. Recent large-scale analysis of over one million preprints and journal articles finds that LLM-modified content appears in up to 22\% of computer science papers by late 2024, with adoption rates rising steadily across disciplines~\cite{liang_2025_quantifying}. Researchers now routinely use systems such as Claude, Gemini, and ChatGPT to locate papers, retrieve bibliographic metadata, draft literature reviews, generate citation lists, and identify research directions~\cite{luo_2025_llm4sr, scherbakov_2025_the, binz_2025_how}. However, scholarly work requires a level of factual precision that differs from general-purpose language generation. A fabricated DOI, an incorrect citation, or a plausible but unverified research direction can propagate through manuscripts, databases, and reviews, creating errors that are difficult to detect and costly to correct~\cite{topaz_2026_fabricated}.

The key failure mode examined in this study is overconfidence. Unlike hallucination in the narrow sense of outright fabrication, overconfidence refers to the tendency of models to produce complete, assertive, and well-formatted outputs even when the required knowledge is incomplete or unreliable. We adopt this term in a deliberately behavioral sense: we treat overconfidence as an observable property of model outputs---assertive, authoritative presentation under conditions where the content is in fact incorrect or unsupported---rather than in the calibration sense, which compares a model's explicitly elicited confidence against its empirical accuracy. Because our benchmark does not prompt models to report confidence scores, we do not claim to measure calibration error directly, and we treat the relationship between self-reported confidence and accuracy as an important complementary direction for future work. 

Prior studies have documented this overconfidence problem in academic writing. For instance, \citet{alkaissi_2023_artificial} show that ChatGPT can generate structurally valid but fictitious references, \citet{walters_2023_fabrication} quantify fabrication and attribution errors across generated bibliographies, and \citet{mugaanyi_2023_citations} evaluate LLM citation reliability across multiple disciplines, finding consistent accuracy limitations in reference generation. Broader studies of LLM calibration further show that models have limited awareness of their own uncertainty boundaries~\cite{kadavath_2022_language}, and that prompting models to express uncertainty does not reliably improve calibration across knowledge-intensive tasks~\cite{xiong_2023_can}. Despite this evidence, systematic evaluation of overconfidence across multiple academic workflow tasks remains limited, particularly under real-world user-facing conditions.

GIScience provides a suitable domain for evaluating this problem. The field has a well-defined journal landscape, established bibliometric structure, strict metadata conventions, and dense citation networks across subfields~\cite{biljecki_2016_a, juhsz_2024_assessing}. These characteristics make it possible to construct objective ground truth for metadata retrieval, citation linking, and research direction generation using Scopus records. Some existing LLM benchmarks, such as HELM~\cite{liang_2022_holistic} and SciEval~\cite{sun_2024_scieval}, have attempted to tackle this challenge, but they mainly focus on factual question answering or domain knowledge recall under controlled settings, rather than open-ended scholarly workflow tasks under real user-facing conditions.

To address these gaps, we introduce GIScholarBench, a large-scale benchmark for evaluating LLM academic capabilities and overconfidence in GIScience. The benchmark is constructed from 10,865 papers collected from 25 core GIScience journals published between 2020 and 2025. It includes three tasks of increasing cognitive complexity: metadata retrieval, literature linking, and research direction generation. We evaluate Claude Sonnet 4.5, Gemini 3, and ChatGPT 5.3 through their native web interfaces using an automated browser-based collection workflow, ensuring that the observed behavior reflects real-world user-facing conditions rather than controlled API environments.

Based on prior evidence on LLM overconfidence and the increasing cognitive demands of the three tasks, we propose and examine three hypotheses, each corresponding to one task. First, in metadata retrieval, models are expected to produce definitive answers rather than abstain, even when the correct metadata is uncertain. Verification instructions may reduce some errors, but may also encourage models to rewrite outputs for surface consistency rather than rejecting unreliable answers. Second, in literature linking, models are expected to identify one or a few plausible references but struggle to maintain accuracy across a full citation list once reliable retrieval capacity is exhausted. Third, in research direction generation, models are expected to concentrate on mainstream topic clusters and underrepresent frontier or cross-domain directions. Examining these hypotheses is expected to demonstrate that LLM overconfidence is not task-specific but exists across academic workflows, while its form shifts from factual, to relational, to generative overconfidence as task complexity increases.

The contributions of this work are three-fold. First, it introduces a GIScience-focused benchmark for evaluating LLM performance and overconfidence across multiple academic workflow tasks. Second, it provides a systematic evaluation of the performance and overconfidence of deployed LLM systems under realistic user-facing conditions using a web-interface-based approach. Third, it characterizes overconfidence as a task-invariant but form-varying phenomenon, ranging from factual errors in metadata retrieval, to relational errors in citation linking and conservative and mainstream-biased outputs in research direction generation.

\section{Related Work}

\subsection{LLM Hallucination and Overconfidence in Knowledge-Intensive Tasks}

The tendency of LLMs to generate plausible but factually incorrect outputs, commonly referred to as hallucination, has received substantial research attention. Prior studies distinguish between factual hallucinations, which contradict verifiable knowledge, and faithfulness hallucinations, which diverge from provided context or source material~\cite{huang_2024_a}. Beyond hallucination itself, recent work has identified overconfidence as a related but distinct issue. Rather than simply fabricating information, overconfident models continue to produce authoritative and highly certain responses even when operating beyond their reliable knowledge boundary. \citet{kadavath_2022_language} show that LLMs possess limited and inconsistent awareness of their own uncertainty, while \citet{xiong_2023_can} demonstrate that explicitly prompting models to express uncertainty does not reliably improve calibration across knowledge-intensive tasks.

These problems become particularly critical in academic and scientific workflows. \citet{alkaissi_2023_artificial} report that ChatGPT frequently generates fabricated academic references that appear structurally valid despite being nonexistent. \citet{walters_2023_fabrication} further quantify citation fabrication and attribution errors across large-scale generated bibliographies, demonstrating that AI-assisted citation generation remains unreliable without manual verification. More broadly, factual evaluation frameworks such as TruthfulQA~\cite{lin_2022_truthfulqa} and FActScore~\cite{min_2023_factscore} introduce systematic methods for measuring factual correctness in open-ended generation. However, these benchmarks primarily target general-domain factuality rather than scholarly literature tasks. In contrast, this study focuses on overconfidence within academic workflows, particularly metadata retrieval, citation generation, and future research direction prediction in GIScience literature.

\subsection{Benchmarking LLMs for Scientific Literature and Domain Knowledge}

Domain-specific benchmarks have become increasingly important for evaluating whether LLMs satisfy the precision requirements of professional and scientific fields. Existing benchmarks in medicine, law, and science---including MedQA~\cite{jin_2021_what}, LegalBench~\cite{guha_2023_legalbench}, and SciEval~\cite{sun_2024_scieval}---primarily evaluate question answering, reasoning, and domain knowledge understanding. Broader frameworks such as HELM~\cite{liang_2022_holistic} further assess calibration, robustness, and factual accuracy across multiple tasks. Although these benchmarks provide valuable evaluations of domain knowledge, they rarely examine open-ended academic workflow tasks such as literature retrieval, citation linking, or research direction generation.

Within geospatial and GIScience research, prior studies have explored the ability of foundation models and LLMs to support spatial reasoning and geographic knowledge tasks. GeoLLM~\cite{manvi_2023_geollm} and StreetviewLLM~\cite{li_2024_streetviewllm} demonstrate that LLMs encode substantial geospatial knowledge and can be adapted for location-based prediction tasks; \citet{huang_2025_evaluating} further benchmark LLMs on geometric classification, topological relations, and direction estimation tasks, revealing systematic gaps in spatial reasoning, while \citet{mai_2024_on} review the broader opportunities and challenges of foundation models in geospatial AI. Bibliometric studies have also analyzed the publication structure and journal landscape of GIScience~\cite{biljecki_2016_a, juhsz_2024_assessing}, providing the disciplinary basis for corpus construction in this work. However, to the best of our knowledge, no existing benchmark has systematically evaluated LLM's performance and overconfidence behavior within GIScience scholarly workflows, particularly under real-world user-facing web interfaces rather than controlled API environments. GIScholarBench is poised to address this gap by constructing a large-scale benchmark that evaluates metadata retrieval, literature linking, and future research direction generation across deployed LLM systems.

\section{Methodology}

\begin{figure*}[htbp]
  \centering
  \includegraphics[width=\textwidth]{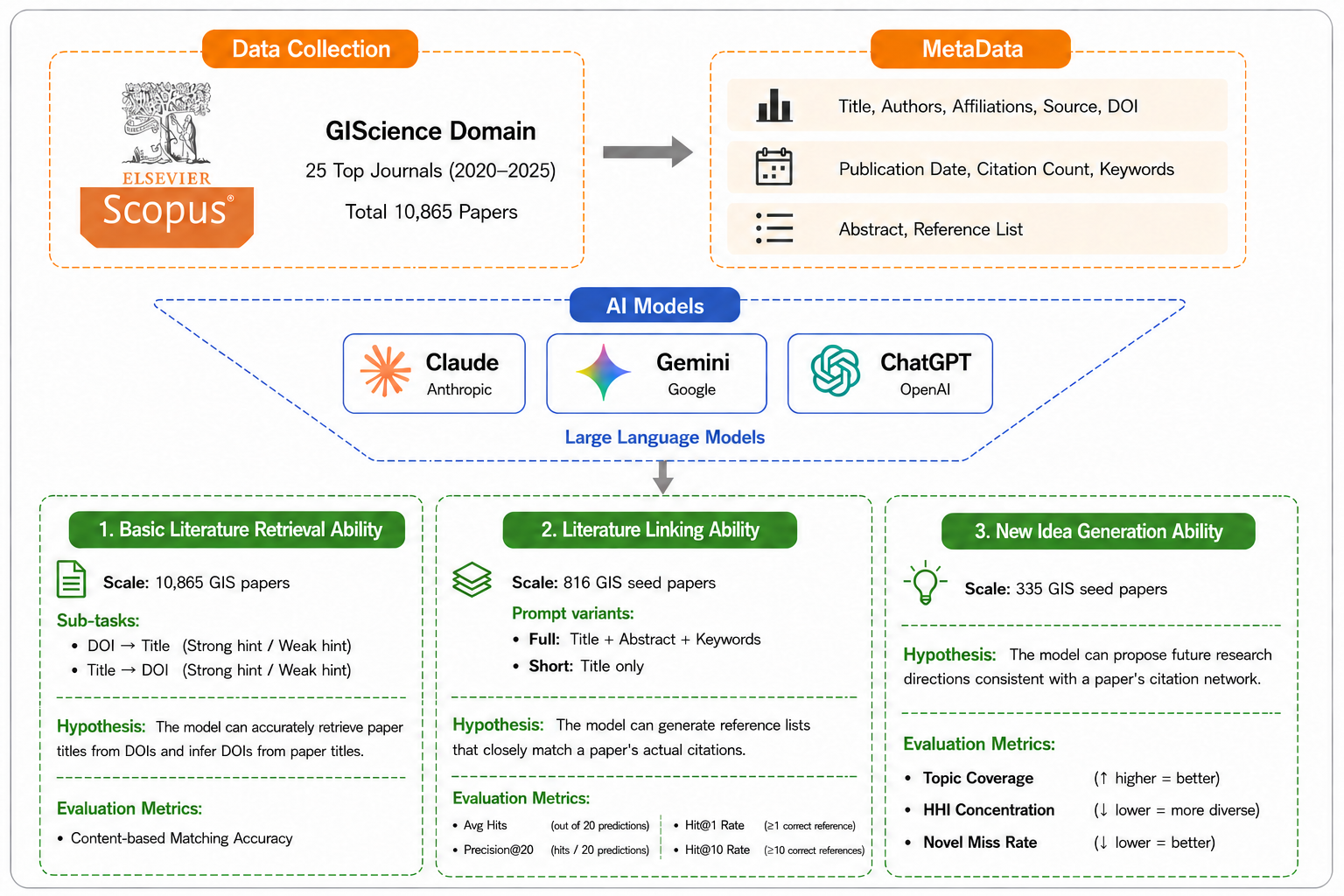}
  \caption{Overview of the GIScholarBench Construction Pipeline and Evaluation Framework.}
  \label{fig:pipeline}
\end{figure*}

\subsection{Dataset Construction}

We construct the benchmark dataset from publications indexed in the Scopus database. To ensure domain relevance and disciplinary consistency, we identify 25 core GIScience journals based on prior bibliometric surveys of the GIScience literature~\cite{biljecki_2016_a, juhsz_2024_assessing}. Journals are retained if they satisfy three criteria: (1) continuous indexing in Scopus during the study period, (2) a primary focus on geographic information science, spatial analysis, remote sensing--GIS integration, or spatial cognition, and (3) uninterrupted publication activity between 2020 and 2025. The complete journal list and abbreviations are provided in Table~\ref{tab:journals}.

We retrieve publications published between January 2020 and December 2025 through the Scopus API. After removing duplicate records and entries without a resolvable DOI or abstract, the final corpus contains 10,865 articles. For each article, we collect ten structured metadata fields: title, author list, author affiliations, source journal, DOI, publication year, citation count, author-assigned keywords, abstract, and reference list. These metadata fields provide the basis for evaluating different dimensions of academic intelligence, particularly literature linking and research direction generation, which rely on both semantic and citation-based contextual information.

\begin{figure*}[htbp]
  \centering
  \includegraphics[width=\textwidth, trim=0 0 0 60pt, clip]{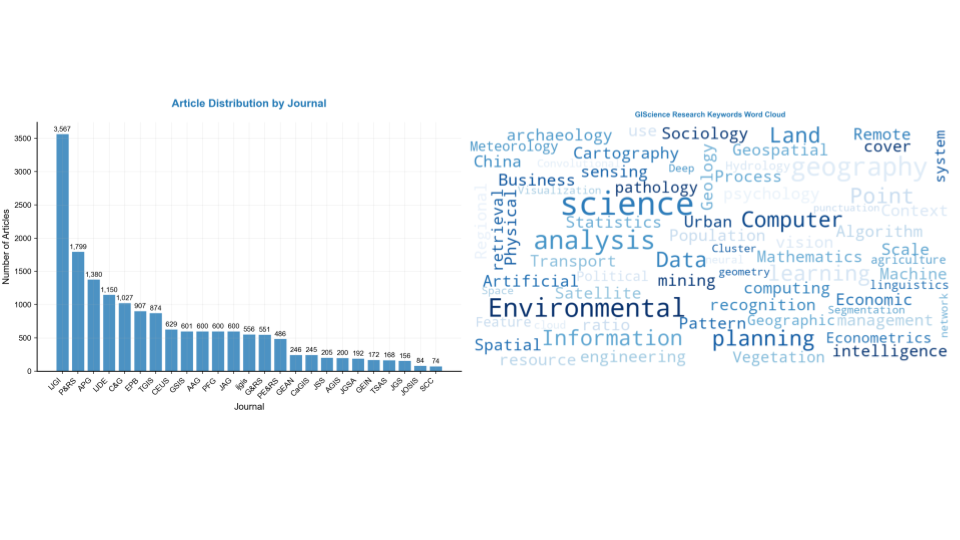}
  \caption{Distribution of Articles and Research Keywords Across the GIScience Benchmark Dataset.}
  \label{fig:distribution}
\end{figure*}

The resulting corpus exhibits substantial disciplinary diversity and publication imbalance across journals, reflecting the heterogeneous nature of GIScience research. As illustrated in Figure~\ref{fig:distribution}, journals such as the \textit{International Journal of Geographical Information Science} and the \textit{ISPRS Journal of Photogrammetry and Remote Sensing} contribute a large proportion of the collected articles, while smaller domain-specific venues provide complementary coverage of specialized research topics. The keyword distribution further demonstrates the interdisciplinary characteristics of the dataset, spanning themes related to geographic information science, environmental studies, spatial analysis, machine learning, remote sensing, urban planning, and computational methods.

\begin{table*}[htbp]
\caption{Core GIScience Journals Included in the Benchmark Dataset}
\label{tab:journals}
\small
\begin{tabular}{ll|ll}
\toprule
\textbf{Abbrev.} & \textbf{Full Name} & \textbf{Abbrev.} & \textbf{Full Name} \\
\midrule
AAG   & Annals of the Association of American Geographers         & IJGI  & ISPRS International Journal of Geo-Information \\
AGIS  & Annals of GIS                                            & JAG   & International Journal of Applied Earth Observation \\
APG   & Applied Geography                                        & JGS   & Journal of Geographical Systems \\
CaGIS & Cartography and Geographic Information Science           & JGSA  & Journal of Geovisualization and Spatial Analysis \\
C\&G  & Computers \& Geosciences                                 & JOSIS & Journal of Spatial Information Science \\
CEUS  & Computers, Environment and Urban Systems                 & JSS   & Journal of Spatial Science \\
EPB   & Environment and Planning B: Urban Analytics and City Science & P\&RS & ISPRS Journal of Photogrammetry and Remote Sensing \\
GEAN  & Geographical Analysis                                    & PE\&RS & Photogrammetric Engineering \& Remote Sensing \\
GEIN  & Geoinformatica                                          & PFG   & Photogrammetrie, Fernerkundung, Geoinformation \\
G\&RS & GIScience \& Remote Sensing                              & SCC   & Spatial Cognition \& Computation \\
GSIS  & Geo-spatial Information Science                         & TGIS  & Transactions in GIS \\
IJDE  & International Journal of Digital Earth                  & TSAS  & ACM Transactions on Spatial Algorithms and Systems \\
IJGIS & International Journal of Geographical Information Science & & \\
\bottomrule
\end{tabular}
\end{table*}

\subsection{Task Design and Evaluation}

\subsubsection{Task 1: Metadata Retrieval}

We design Task 1 to evaluate whether LLMs can accurately retrieve fundamental bibliographic metadata for GIScience publications. The task includes two retrieval directions: retrieving the article title from a DOI (DOI$\rightarrow$Title) and retrieving the DOI from an article title (Title$\rightarrow$DOI). Each direction is further evaluated under two prompt conditions, Strong and Weak, resulting in four sub-tasks in total.

For the DOI$\rightarrow$Title task, the Strong condition provides the model with the article DOI, abstract, and keywords, together with an explicit instruction to verify that the retrieved title is semantically consistent with the supplied abstract. The Weak condition removes the verification instruction and only asks the model to retrieve the corresponding title from bibliographic metadata. For the Title$\rightarrow$DOI task, the Strong condition includes the article title, abstract, and keywords, whereas the Weak condition reduces the input to the title and a coarse domain descriptor (``GIScience''). This design enables us to isolate the influence of contextual richness and instruction specificity on retrieval performance. Full prompt templates for all four sub-tasks are provided in the Appendix (Figure~\ref{fig:promptA1}).

Ground-truth answer pairs are directly constructed from the Scopus metadata collected in Section~3.1. The canonical title and DOI associated with each of the 10,865 articles serve as the reference labels, requiring no additional manual annotation.

For the DOI$\rightarrow$Title task, we compare the model-generated title $\hat{t}$ with the ground-truth title $t^{*}$ after basic text normalization. Let $K(\cdot)$ denote the set of content words after stopword removal. A prediction is considered correct if:

\begin{equation}
\text{correct} = \begin{cases}
1 & \text{if } \hat{t} = t^{*} \\
1 & \text{if } t^{*} \subseteq \hat{t} \text{ or } \hat{t} \subseteq t^{*} \\
1 & \text{if } \dfrac{|K(\hat{t}) \cap K(t^{*})|}{|K(t^{*})|} \geq 0.6 \\
0 & \text{otherwise}
\end{cases}
\end{equation}

For the Title$\rightarrow$DOI task, let $D(\hat{r})$ represent the set of DOI strings extracted from the model response $\hat{r}$ using a regular expression pattern designed to identify standard DOI formats. A prediction is considered correct if the ground-truth DOI $d^{*}$ appears in the extracted DOI set:

\begin{equation}
\text{correct} = \mathbf{1}[d^{*} \in D(\hat{r})]
\end{equation}

For both retrieval directions, we compute overall accuracy as:

\begin{equation}
\text{Acc} = \frac{1}{N}\sum_{i=1}^{N} \mathbf{1}[\text{correct}_{i}], \quad N = 10{,}865
\end{equation}

where $N$ is fixed to the full corpus size regardless of the number of successfully collected responses, thereby proportionally penalizing incomplete response generation or failed retrieval attempts.

\subsubsection{Task 2: Literature Linking}

We design Task 2 to evaluate whether large language models (LLMs) can recover the citation-network neighborhood of a given GIScience paper by identifying related academic references. Specifically, models are asked to generate up to 20 real and verifiable papers that are likely to be cited by the seed paper or to cite the seed paper. Two prompt variants are evaluated. The Full setting provides the seed paper's title, abstract, and keywords, whereas the Short setting includes only the title. This design enables us to examine how additional bibliographic context influences citation-link prediction performance. Full prompt templates are provided in the Appendix (Figure~\ref{fig:promptA2}).

Ground-truth citation sets are constructed from Scopus citation records for each of the 816 seed papers. For every seed paper, we collect all papers within the benchmark corpus that either cite the seed paper or are cited by it. The resulting citation neighborhoods range from 1 to 352 references, with an average of 47.3 references per seed paper. Note that ground-truth coverage is bounded by the 10,865-paper corpus; references outside this set are not included, making the reported hit metrics conservative lower-bound estimates of true model recall.

Because model-generated titles may not exactly match the wording of ground-truth references, we adopt a fuzzy title-matching strategy based on TF-IDF cosine similarity. For each predicted title $\hat{s}$, we construct a hybrid feature representation combining word-level TF-IDF features with character-level $n$-gram TF-IDF features ($n = 3$--$5$). The similarity between a predicted title and a ground-truth title $g$ is computed as:

\begin{equation}
\text{sim}(\hat{s},\, g) = \cos\!\left(\phi(\hat{s}),\, \phi(g)\right), \quad
\phi(\cdot) = \bigl[\phi_{\text{word}}(\cdot)\,\|\,\phi_{\text{char}}(\cdot)\bigr]
\end{equation}

A prediction is considered successfully matched if the similarity score exceeds a threshold of $\tau = 0.20$. Each ground-truth reference can only be matched once to avoid duplicate counting. Let $M_i$ denote the number of successfully matched references for seed paper $i$, and $\hat{n}_i$ denote the number of predicted references. We report four evaluation metrics over the set $\mathcal{Q}$ of 816 seed papers:

\begin{equation}
\text{Avg Hits} = \frac{1}{|\mathcal{Q}|}\sum_{i \in \mathcal{Q}} M_i
\end{equation}

\begin{equation}
\text{P@20} = \frac{1}{|\mathcal{Q}|}\sum_{i \in \mathcal{Q}} \frac{M_i}{\min(\hat{n}_i,\;20)}
\end{equation}

\begin{equation}
\text{Hit@}k = \frac{1}{|\mathcal{Q}|}\sum_{i \in \mathcal{Q}} \mathbf{1}[M_i \geq k], \quad k \in \{1,\,10\}
\end{equation}

These metrics jointly evaluate the ability of LLMs to recover relevant citation relationships, while also measuring the precision and consistency of the generated literature links.

\subsubsection{Task 3: Research Direction Generation}

\begin{figure*}[htbp]
  \centering
  \includegraphics[width=\textwidth]{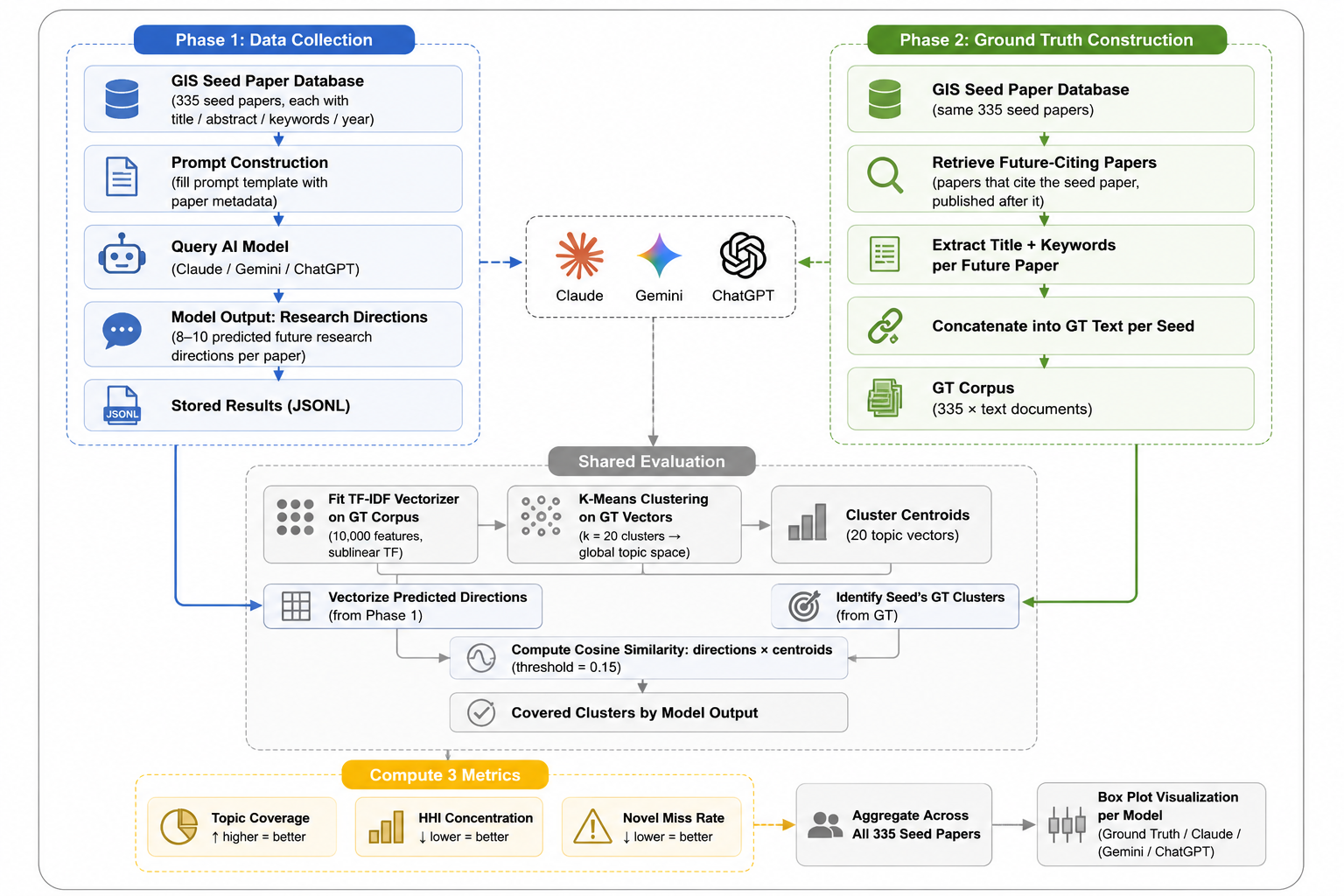}
  \caption{Overall Evaluation Framework for Future Research Direction Generation and Topic-Space Assessment.}
  \label{fig:framework}
\end{figure*}

We design Task 3 to evaluate whether large language models (LLMs) can anticipate future research directions emerging from existing GIScience studies, following the idea of using subsequent scientific impact to assess the breadth of AI-generated research directions~\cite{hao_2026_artificial}. Specifically, models are asked to generate 8--10 concrete and distinct research directions that could naturally extend a given seed paper. Each prompt includes the seed paper's title, journal, publication year, citation count, keywords, and abstract. Models are instructed to produce actionable and diverse directions spanning different methods, spatial scales, regions, and research problems while avoiding simple restatements of the original paper's conclusions. The full prompt template is provided in the Appendix (Figure~\ref{fig:promptA2}, bottom), and the overall evaluation framework is illustrated in Figure~\ref{fig:framework}.

Ground-truth future research directions are constructed from subsequent papers that cite each seed paper. For each of the 335 seed papers, we collect all future-citing papers published after the seed paper's publication year. The titles and keywords of these future-citing papers are concatenated into a single ground-truth document representing the actual research trajectories later pursued by the GIScience community.

Because both generated directions and ground-truth references are represented as free-form natural language, direct string matching is not appropriate. Instead, we project both into a shared topic space derived from the ground-truth corpus. We first fit a TF-IDF vectorizer with 10,000 features and sublinear term-frequency scaling on all ground-truth documents. We then apply K-Means clustering ($k = 20$) to derive 20 topic centroids $\{c_1, \ldots, c_{20}\}$ representing the global landscape of future GIScience research directions. The value $k = 20$ was selected based on the elbow method applied to within-cluster sum of squares, and produces clusters that are interpretable as distinct GIScience research themes without over-fragmenting the topic space.

For each seed paper $i$, the ground-truth topic coverage set is defined as:
\begin{equation}
C_i^{\text{GT}} = \bigl\{j \in \{1,\ldots,k\} : \exists\, g \in G_i,\;
\cos\!\left(\phi(g),\, c_j\right) \geq \delta \bigr\}
\end{equation}
where $\phi(\cdot)$ denotes the TF-IDF embedding function and $\delta = 0.15$ is the cluster assignment threshold. Similarly, the predicted topic coverage set is:
\begin{equation}
C_i^{\text{pred}} = \bigl\{j \in \{1,\ldots,k\} : \exists\, d \in D_i,\;
\cos\!\left(\phi(d),\, c_j\right) \geq \delta \bigr\}
\end{equation}

We evaluate model performance using three complementary metrics. Topic Coverage (TC) measures the proportion of ground-truth research topics successfully recovered by the generated directions:
\begin{equation}
\text{TC}_i = \frac{|C_i^{\text{pred}} \cap C_i^{\text{GT}}|}{|C_i^{\text{GT}}|}, \qquad
\overline{\text{TC}} = \frac{1}{|\mathcal{Q}|}\sum_{i \in \mathcal{Q}} \text{TC}_i
\end{equation}

Novel Miss Rate (NMR) measures the proportion of ground-truth research directions not captured by the model outputs:
\begin{equation}
\text{NMR}_i = \frac{|C_i^{\text{GT}} \setminus C_i^{\text{pred}}|}{|C_i^{\text{GT}}|}, \qquad
\overline{\text{NMR}} = \frac{1}{|\mathcal{Q}|}\sum_{i \in \mathcal{Q}} \text{NMR}_i
\end{equation}

To evaluate thematic diversity, we compute the Herfindahl--Hirschman Index (HHI), which measures the concentration of generated directions across topic clusters:
\begin{equation}
\text{HHI}_i = \sum_{j=1}^{k}\!\left(\frac{n_{ij}}{|D_i|}\right)^{\!2}, \qquad
\overline{\text{HHI}} = \frac{1}{|\mathcal{Q}|}\sum_{i \in \mathcal{Q}} \text{HHI}_i
\end{equation}
where $n_{ij}$ denotes the number of generated directions assigned to cluster $j$, and $|D_i|$ denotes the total number of generated directions for seed paper $i$. Higher Topic Coverage and lower Novel Miss Rate and HHI values indicate stronger performance in anticipating diverse and realistic future research trajectories.

\subsection{Data Collection Strategy}

We collect model responses through each platform's native web interface rather than through official APIs. This strategy better reflects the environment used by most real-world users and captures product-level behaviors such as retrieval augmentation, knowledge-cutoff handling, response formatting, and refusal patterns. It also avoids potential differences between API-based and web-based model deployments. The three evaluated systems are Claude Sonnet 4.5 (Anthropic), Gemini 3 (Google), and ChatGPT 5.3 (OpenAI), accessed through their respective web chat interfaces.

\begin{figure}[htbp]
  \centering
  \includegraphics[width=\columnwidth]{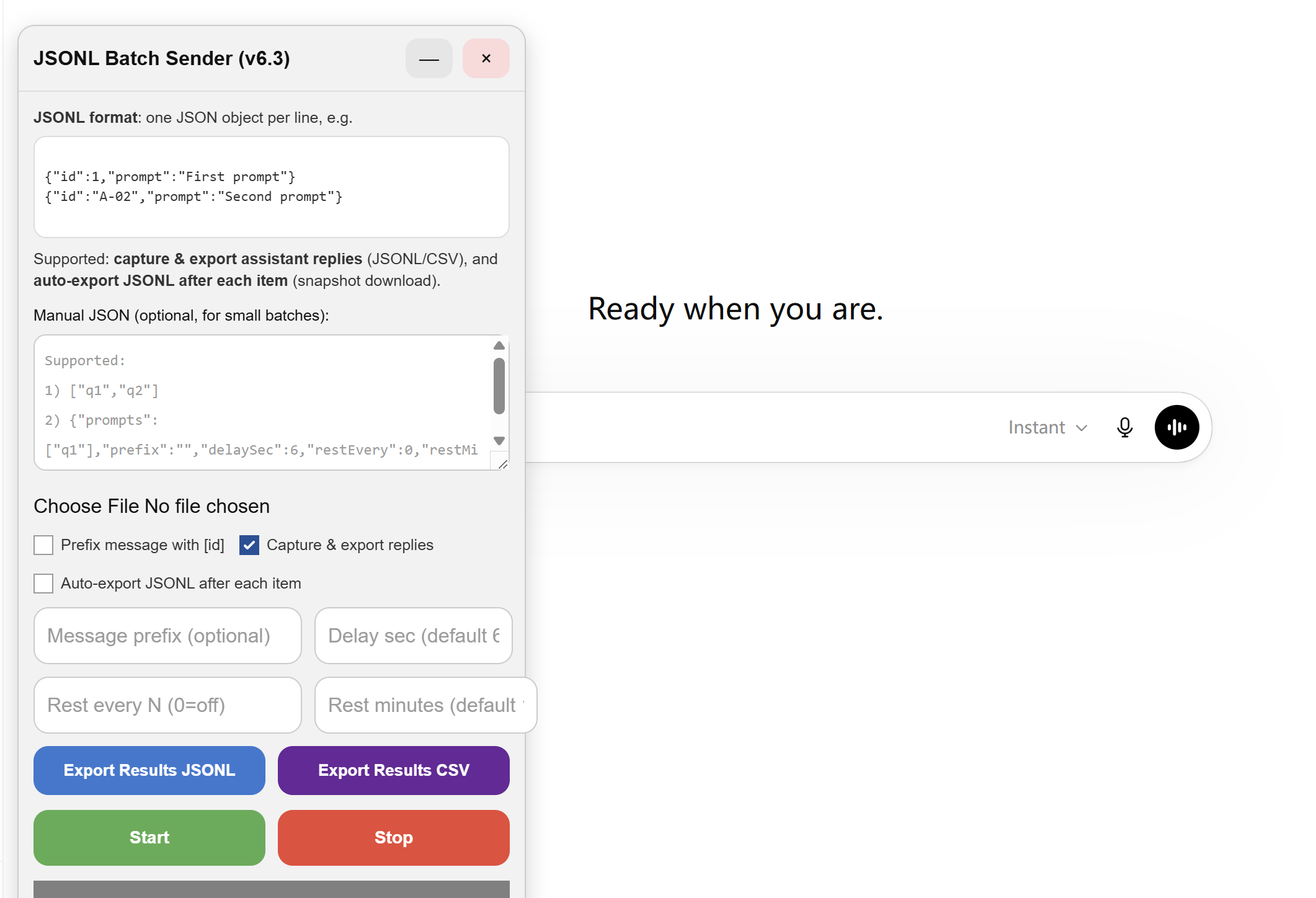}
  \caption{Automated Web-Based Prompt Collection Using the JSONL Batch Sender Extension.}
  \label{fig:collection}
\end{figure}

To automate response collection at scale, we use the JSONL Batch Sender (v6.3) browser extension (Figure~\ref{fig:collection}). For each task and model, we prepare a JSONL prompt file in which each line contains a unique item identifier and a fully instantiated prompt. The extension loads this file, submits prompts sequentially through the chat interface, waits for each model response to complete, and records the generated output text.

Responses are exported incrementally in JSONL format after each query, which provides fault-tolerant snapshots and allows interrupted sessions to be resumed. To reduce the risk of exceeding platform limits or triggering anti-automation mechanisms, we apply configurable delays between messages and periodic rest intervals during collection. This workflow enables reproducible large-scale response collection across all three tasks and three models using only their public web interfaces.

\section{Results}

\subsection{Metadata Retrieval Performance}

\begin{figure*}[htbp]
  \centering
  \includegraphics[width=\textwidth]{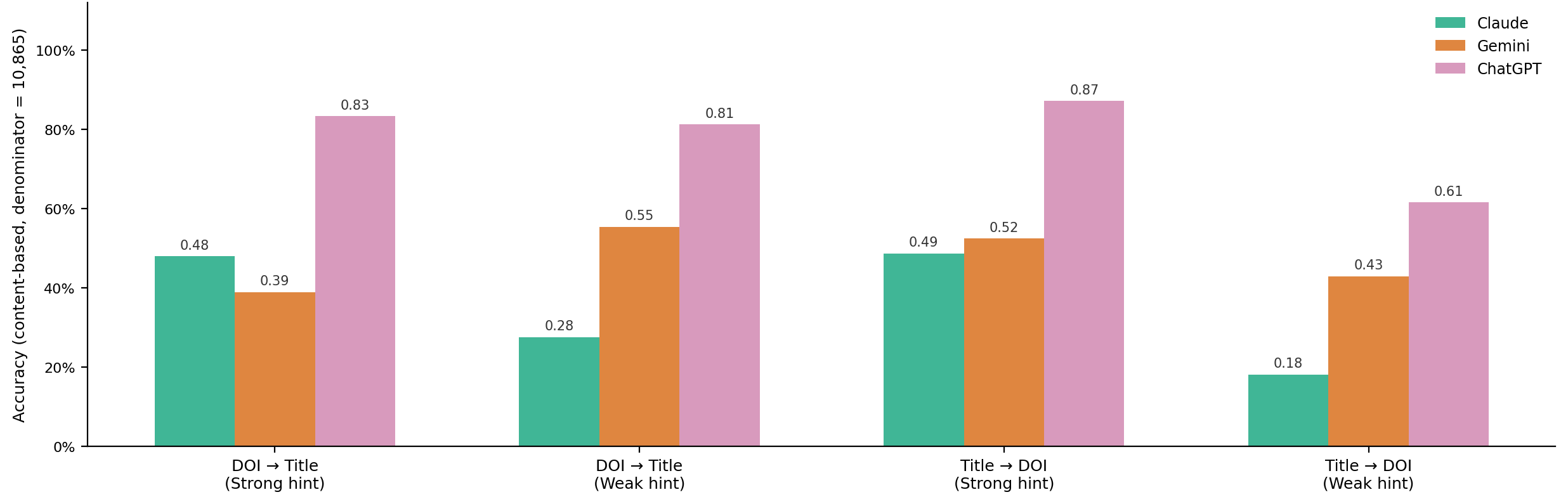}
  \caption{Metadata Retrieval Accuracy Across Task~1 Settings (content-based matching, denominator = 10,865).}
  \label{fig:task1}
\end{figure*}

Figure~\ref{fig:task1} presents the metadata retrieval accuracy across four Task~1 settings. The central finding is not only that models differ in accuracy, but that all three models tend to produce complete and confident bibliographic answers even when those answers are incorrect. ChatGPT achieves the highest accuracy overall, reaching 0.87 in the Title$\rightarrow$DOI Strong setting, while Claude and Gemini perform lower in several settings. However, none of the models meaningfully signal uncertainty or abstain from answering when retrieval fails. This suggests that metadata retrieval is often treated as a generative task, where models produce plausible titles or DOIs rather than explicitly recognizing the limits of their knowledge.

The Strong and Weak settings further reveal how overconfidence appears under different forms of context. In the Title$\rightarrow$DOI task, additional context improves accuracy for all models, indicating that abstracts and keywords help disambiguate publications. In contrast, the DOI$\rightarrow$Title task shows weaker or inconsistent gains: Claude decreases from 0.48 to 0.28, ChatGPT remains nearly unchanged, and only Gemini improves. In these cases, models often generate titles that are semantically compatible with the supplied abstract but do not match the actual paper title. This behavior reflects a key form of overconfidence in scholarly metadata retrieval: models do not simply fail silently, but often present plausible, well-formed, and authoritative answers that mask factual errors.

\subsection{Citation Linking Performance}

\begin{figure*}[htbp]
  \centering
  \includegraphics[width=\textwidth]{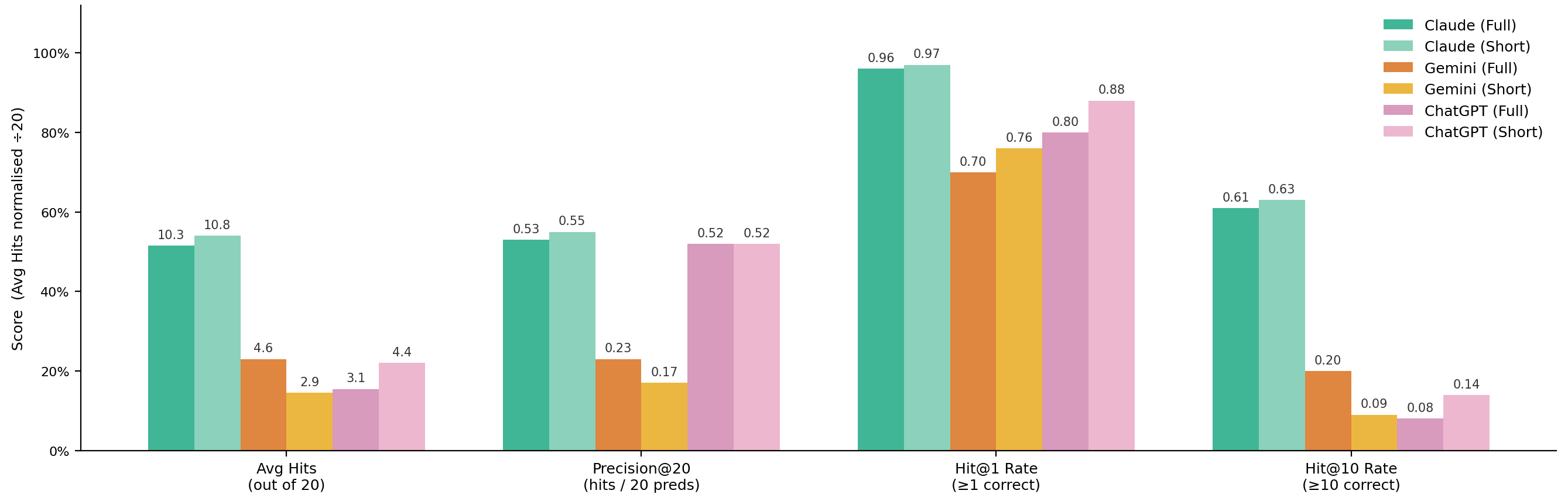}
  \caption{Literature Linking Performance Across Models and Prompt Conditions (content-based matching).}
  \label{fig:task2}
\end{figure*}

Figure~\ref{fig:task2} summarizes the Task~2 results across Avg Hits, P@20, Hit@1, and Hit@10. A consistent pattern emerges across all models and prompt settings: models are generally able to retrieve at least one relevant citation for most seed papers, but performance declines substantially when generating longer citation lists. This gap between partial retrieval success and overall citation accuracy reveals a key characteristic of LLM-based literature linking: models often continue generating plausible references even after reliable retrieval capacity has been exhausted.

The contrast between Hit@1 and Hit@10 is particularly notable. Claude achieves the strongest overall performance, with Hit@1 rates above 0.96 and Hit@10 rates above 0.60 in both prompt settings. Gemini and ChatGPT also achieve relatively strong Hit@1 performance, indicating that the models can usually identify at least one topically relevant citation. However, Hit@10 drops sharply for these models, reaching only 0.20 for Gemini Full, 0.08 for ChatGPT Full, and 0.14 for ChatGPT Short. Precision@20 further highlights this issue. Gemini achieves P@20 values of only 0.23 and 0.17, implying that most generated references cannot be matched to verified citation records. ChatGPT achieves higher precision values (approximately 0.52), partly because it tends to generate shorter reference lists rather than fully populating all 20 entries. Even Claude, despite obtaining the highest Avg Hits and Hit@10 performance, still produces a substantial proportion of unmatched citations. Across all evaluated settings, no model exceeds a P@20 value of 0.55, indicating that a large fraction of generated references remain unverifiable despite being presented in fluent and authoritative formats.

\subsection{Research Direction Generation Performance}

\begin{figure*}[htbp]
  \centering
  \includegraphics[width=\textwidth]{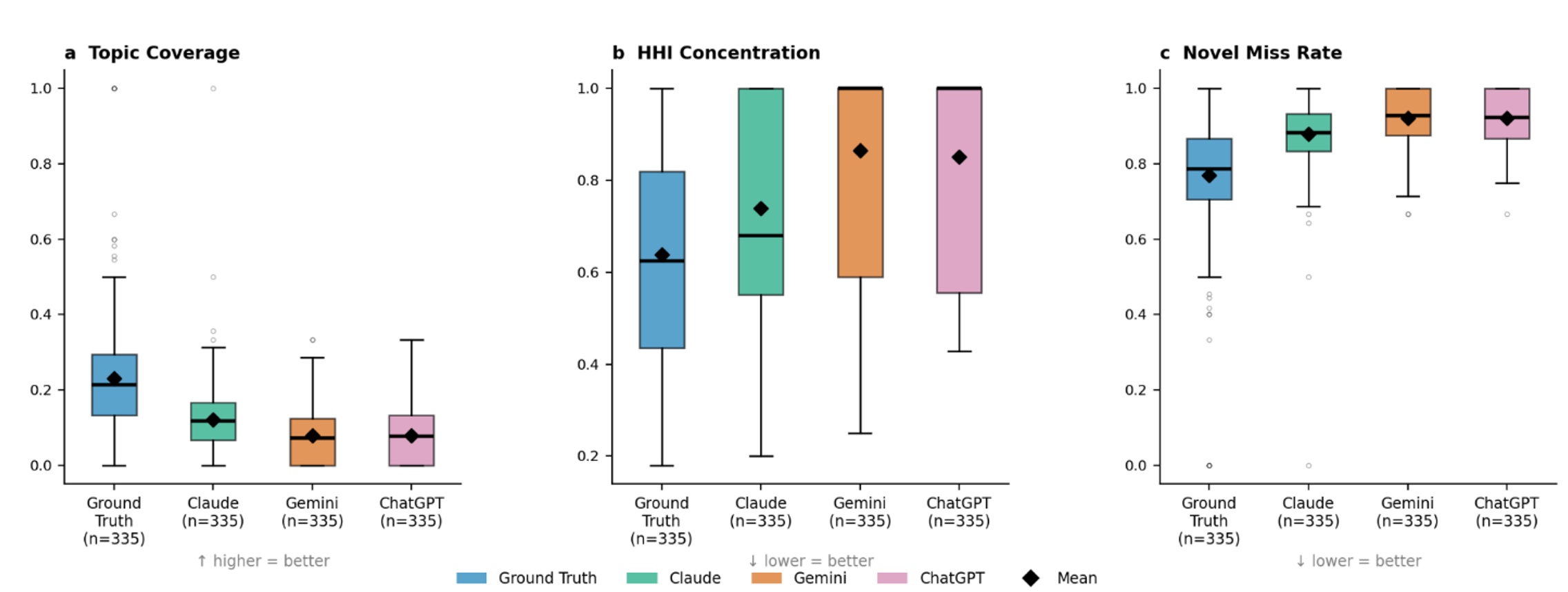}
  \caption{Comparative Evaluation of AI-Generated and Ground-Truth Research Directions in Task~3 (box plots with mean diamonds; $n=335$ for all groups).}
  \label{fig:task3}
\end{figure*}

\begin{figure*}[htbp]
  \centering
  \includegraphics[width=0.72\textwidth]{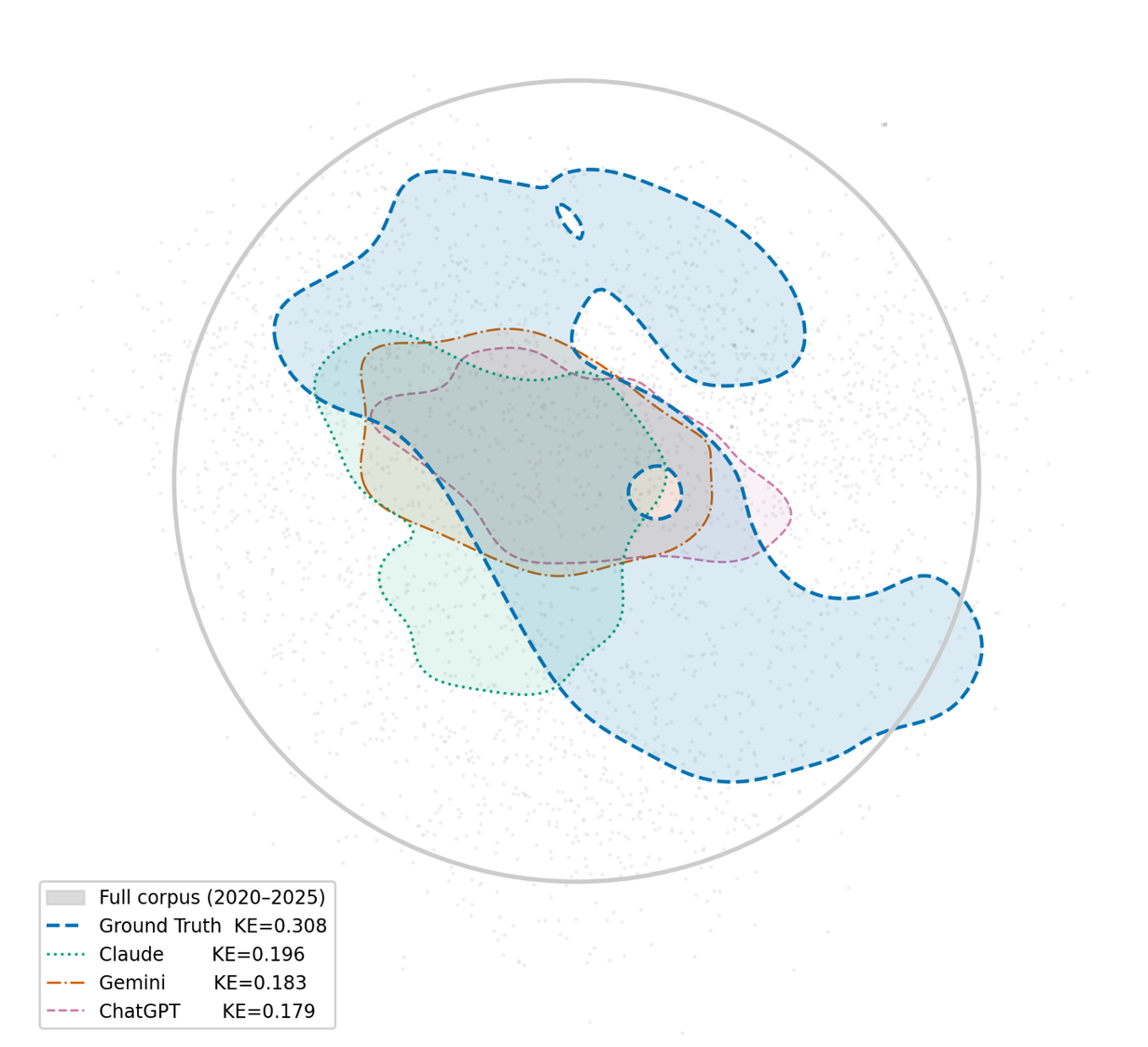}
  \caption{Distribution of Ground-Truth and AI-Generated Research Directions in GIS Semantic Space (kernel entropy KE values shown in legend).}
  \label{fig:task3kde}
\end{figure*}

Figures~\ref{fig:task3} and~\ref{fig:task3kde} summarize the Task~3 results from both metric-based and distributional perspectives. Across all models, generated research directions exhibit substantially narrower topical coverage than the trajectories observed in real future-citing literature. Although the generated directions are typically presented as concrete and actionable research ideas, the evaluation results indicate that the models cover only a limited portion of the broader GIScience research space.

As shown in Figure~\ref{fig:task3}(a), the Ground Truth baseline achieves a mean Topic Coverage of approximately 0.22, whereas Claude reaches 0.12 and both Gemini and ChatGPT remain near 0.09. The corresponding Novel Miss Rate results in Figure~\ref{fig:task3}(c) show that Gemini and ChatGPT fail to capture approximately 91\% of the topic space represented by real downstream research. Figure~\ref{fig:task3}(b) further demonstrates that model-generated directions are substantially more concentrated than real research trajectories. The Ground Truth baseline achieves a mean HHI of approximately 0.64, while Claude increases to 0.70 and both Gemini and ChatGPT approach 0.87, indicating that many generated direction lists collapse into a small number of highly similar topic clusters rather than spanning diverse research themes.

Figure~\ref{fig:task3kde} reveals a similar pattern at the distributional level. The Ground Truth occupies a broad and spatially dispersed region of the GIS semantic space, achieving a kernel entropy (KE) of 0.308. In contrast, all three AI models produce substantially smaller and more centrally concentrated distributions, with KE values of 0.196 for Claude, 0.183 for Gemini, and 0.179 for ChatGPT. The model-generated distributions overlap heavily with one another but cover only a limited portion of the broader semantic space occupied by real future research. Peripheral regions associated with emerging, cross-domain, or less conventional directions remain largely absent from the generated outputs.

\section{Discussion}

\subsection{Overconfidence as a Task-Invariant but Form-Varying Phenomenon}

The results across all three tasks reveal a consistent pattern: overconfidence persists across different academic workflows, but its form changes with task structure. In Task~1, overconfidence appears as factual overgeneration. Models continue producing specific bibliographic metadata even when retrieval accuracy is limited, with little uncertainty or abstention. This is especially clear in the DOI$\rightarrow$Title Strong setting, where added verification instructions do not consistently improve accuracy. Claude declines from 0.48 to 0.28, and ChatGPT remains nearly unchanged from 0.83 to 0.81. Instead of rejecting uncertain answers, models often generate titles that appear semantically aligned with the abstract but differ from the actual paper title. This suggests that models prioritize coherent and complete responses over strict factual fidelity.

In Task~2, overconfidence becomes relational. Models can often identify at least one plausible citation, as shown by high Hit@1 scores ranging from 0.70 to 0.97. However, Hit@10 drops sharply, especially for Gemini Full (0.20) and ChatGPT Full (0.08), indicating that models struggle to recover multiple verified citation links. Precision@20 never exceeds 0.55, meaning that a substantial share of generated references remains unmatched despite fluent formatting. In Task~3, overconfidence takes a meta-cognitive form. Model-generated directions have lower Topic Coverage than actual future-citing papers, with AI models ranging from 0.09 to 0.12 compared with 0.22 for Ground Truth. They also show higher Novel Miss Rate (0.88--0.91), stronger topic concentration (HHI = 0.70--0.87 vs.\ 0.64 for Ground Truth), and lower semantic-space diversity (KE = 0.179--0.196 vs.\ 0.308 for Ground Truth). Critically, the individual directions are not factually wrong; rather, the overconfidence lies in the model's implicit claim of completeness. Models produce 8--10 directions with no hedging about what might be absent, yet cover only a small fraction of the topic space occupied by real future research. This reflects what \citet{kadavath_2022_language} identify as a core limitation of LLM self-knowledge: models do not know what they do not know. In Task~3, this manifests as overconfidence about the coverage and representativeness of the generated output, not about the correctness of individual items.

\subsection{Implications for AI-Assisted GIS Research Practice}

The findings of this study have direct implications for GIScience researchers who use LLMs in scholarly workflows. Because overconfidence appears across all three tasks and all evaluated models, the risk is not limited to poor prompts or isolated failure cases. Even under routine, well-formed queries, models often produce fluent and authoritative outputs that require verification. For metadata retrieval, even the best-performing model still returns incorrect titles or DOIs in a nontrivial share of cases, so LLM-generated metadata should not be directly used in manuscripts, citation records, or database queries without checking authoritative sources such as Scopus, Crossref, or DOI registries. For literature linking, the high Hit@1 scores (0.70--0.97) suggest that LLMs can help users identify initial anchor papers and quickly enter an unfamiliar topic. However, the much lower Hit@10 scores (0.08--0.61) and P@20 values below 0.55 show that complete generated reference lists remain incomplete and often unverifiable. Therefore, AI-generated citations should be treated as preliminary leads rather than reliable bibliographic outputs.

For research direction generation, the risk is less visible because the outputs are usually coherent, specific, and actionable. However, our results show that AI-generated ideas cover a much narrower portion of the GIScience research space than real future-citing papers, with Topic Coverage of only 0.09--0.12 compared with 0.22 for Ground Truth. This indicates that LLMs tend to reproduce familiar and mainstream directions while underrepresenting emerging, interdisciplinary, or unconventional trajectories. More productive AI-assisted GIScience workflows should position LLMs as auxiliary tools for orientation, filtering, and idea structuring, while relying on human researchers to verify factual claims, trace citations systematically, and actively search for directions that the model may overlook.

\subsection{Limitations and Future Directions}

Several limitations of this study should be acknowledged. First, the browser-based collection strategy improves ecological validity by evaluating models through real-world user-facing web interfaces, but it also introduces reproducibility challenges. Web-deployed LLM systems are continuously updated through model revisions, interface changes, and platform-level adjustments that are not always publicly documented. As a result, the behaviors observed in this study reflect the deployed model versions available during the collection period, and future replications using identical prompts may produce different outputs.

Second, the benchmark is constructed entirely from GIScience literature, and the extent to which these findings generalize to other scientific domains remains uncertain. GIScience has a distinctive interdisciplinary structure and relies heavily on domain-specific terminology, spatial reasoning, and geospatial methodologies, which may influence retrieval accuracy and overconfidence behavior differently from fields such as medicine, law, or the social sciences. Future work should therefore extend the benchmark to additional disciplines and model families, including open-weight, domain-fine-tuned, and retrieval-augmented systems. Such extensions would help determine whether overconfidence is a general property of current LLM architectures or whether specialized corpora, retrieval mechanisms, confidence scoring, and explicit ``Not Found'' constraints can improve factual reliability, topic coverage, and calibration in scholarly workflows.

Third, our analysis characterizes overconfidence as a behavioral property of model outputs and does not measure confidence calibration in the strict sense. We do not elicit explicit confidence estimates from the models, so we cannot quantify the gap between a model's self-reported certainty and its empirical accuracy (for example, via reliability diagrams or expected calibration error). Establishing this calibration relationship---and thereby statistically validating overconfidence in the calibration sense---is an important direction that our current design leaves open.

\section{Conclusion}

This paper presents GIScholarBench, a large-scale benchmark for evaluating LLM academic capabilities and overconfidence in GIScience scholarly workflows. Built from 10,865 papers from 25 core GIScience journals published between 2020 and 2025, the benchmark evaluates metadata retrieval, literature linking, and research direction generation across Claude, Gemini, and ChatGPT through native web interfaces. The results show that LLMs often produce fluent and authoritative outputs even when their answers are incorrect, incomplete, or poorly calibrated. In metadata retrieval, models generate specific titles and DOIs despite factual errors. In literature linking, they can identify some relevant citations but continue producing unverifiable references, with Precision@20 never exceeding 0.55. In research direction generation, model outputs concentrate on familiar GIScience themes and show lower topical diversity than real future-citing papers. These findings suggest that LLMs are useful as orientation and brainstorming tools, but should not be treated as authoritative sources for scholarly knowledge. GIScholarBench provides a foundation for evaluating epistemic reliability in AI-assisted GIScience research.

\bibliographystyle{ACM-Reference-Format}
\bibliography{ACM}


\begin{thebibliography}{24}


\ifx \showCODEN    \undefined \def \showCODEN     #1{\unskip}     \fi
\ifx \showISBNx    \undefined \def \showISBNx     #1{\unskip}     \fi
\ifx \showISBNxiii \undefined \def \showISBNxiii  #1{\unskip}     \fi
\ifx \showISSN     \undefined \def \showISSN      #1{\unskip}     \fi
\ifx \showLCCN     \undefined \def \showLCCN      #1{\unskip}     \fi
\ifx \shownote     \undefined \def \shownote      #1{#1}          \fi
\ifx \showarticletitle \undefined \def \showarticletitle #1{#1}   \fi
\ifx \showURL      \undefined \def \showURL       {\relax}        \fi
\providecommand\bibfield[2]{#2}
\providecommand\bibinfo[2]{#2}
\providecommand\natexlab[1]{#1}
\providecommand\showeprint[2][]{arXiv:#2}

\bibitem[Alkaissi and McFarlane(2023)]%
        {alkaissi_2023_artificial}
\bibfield{author}{\bibinfo{person}{Hussam Alkaissi} {and} \bibinfo{person}{Samy
  McFarlane}.} \bibinfo{year}{2023}\natexlab{}.
\newblock \showarticletitle{Artificial hallucinations in ChatGPT: Implications
  in scientific writing}.
\newblock \bibinfo{journal}{\emph{Cureus}}  \bibinfo{volume}{15}
  (\bibinfo{date}{02} \bibinfo{year}{2023}).
\newblock
\href{https://doi.org/10.7759/cureus.35179}{doi:\nolinkurl{10.7759/cureus.35179}}


\bibitem[Biljecki(2016)]%
        {biljecki_2016_a}
\bibfield{author}{\bibinfo{person}{Filip Biljecki}.}
  \bibinfo{year}{2016}\natexlab{}.
\newblock \showarticletitle{A scientometric analysis of selected GIScience
  journals}.
\newblock \bibinfo{journal}{\emph{International Journal of Geographical
  Information Science}}  \bibinfo{volume}{30} (\bibinfo{date}{01}
  \bibinfo{year}{2016}), \bibinfo{pages}{1302--1335}.
\newblock
\href{https://doi.org/10.1080/13658816.2015.1130831}{doi:\nolinkurl{10.1080/13658816.2015.1130831}}


\bibitem[Binz et~al\mbox{.}(2025)]%
        {binz_2025_how}
\bibfield{author}{\bibinfo{person}{Marcel Binz}, \bibinfo{person}{Stephan
  Alaniz}, \bibinfo{person}{Adina Roskies}, \bibinfo{person}{Balazs Aczel},
  \bibinfo{person}{Carl~T. Bergstrom}, \bibinfo{person}{Colin Allen},
  \bibinfo{person}{Daniel Schad}, \bibinfo{person}{Dirk Wulff},
  \bibinfo{person}{Jevin~D. West}, \bibinfo{person}{Qiong Zhang},
  \bibinfo{person}{Richard~M. Shiffrin}, \bibinfo{person}{Samuel~J. Gershman},
  \bibinfo{person}{Vencislav Popov}, \bibinfo{person}{Emily~M. Bender},
  \bibinfo{person}{Marco Marelli}, \bibinfo{person}{Matthew~M. Botvinick},
  \bibinfo{person}{Zeynep Akata}, {and} \bibinfo{person}{Eric Schulz}.}
  \bibinfo{year}{2025}\natexlab{}.
\newblock \showarticletitle{How should the advancement of large language models
  affect the practice of science?}
\newblock \bibinfo{journal}{\emph{Proceedings of the National Academy of
  Sciences}}  \bibinfo{volume}{122} (\bibinfo{date}{01} \bibinfo{year}{2025}).
\newblock
\href{https://doi.org/10.1073/pnas.2401227121}{doi:\nolinkurl{10.1073/pnas.2401227121}}


\bibitem[Guha et~al\mbox{.}(2023)]%
        {guha_2023_legalbench}
\bibfield{author}{\bibinfo{person}{Neel Guha}, \bibinfo{person}{Julian Nyarko},
  \bibinfo{person}{Daniel~E Ho}, \bibinfo{person}{Christopher Ré},
  \bibinfo{person}{Adam Chilton}, \bibinfo{person}{Aditya Narayana},
  \bibinfo{person}{Alex Chohlas-Wood}, \bibinfo{person}{Austin Peters},
  \bibinfo{person}{Brandon Waldon}, \bibinfo{person}{Daniel~N Rockmore},
  \bibinfo{person}{Diego Zambrano}, \bibinfo{person}{Dmitry Talisman},
  \bibinfo{person}{Enam Hoque}, \bibinfo{person}{Faiz Surani},
  \bibinfo{person}{Frank Fagan}, \bibinfo{person}{Galit Sarfaty},
  \bibinfo{person}{Gregory~M Dickinson}, \bibinfo{person}{Haggai Porat},
  \bibinfo{person}{Jason Hegland}, \bibinfo{person}{Jessica Wu},
  \bibinfo{person}{Joe Nudell}, \bibinfo{person}{Joel Niklaus},
  \bibinfo{person}{John Nay}, \bibinfo{person}{Jonathan~H Choi},
  \bibinfo{person}{Kevin Tobia}, \bibinfo{person}{Margaret Hagan},
  \bibinfo{person}{Megan Ma}, \bibinfo{person}{Michael Livermore},
  \bibinfo{person}{Nikon Rasumov-Rahe}, \bibinfo{person}{Nils Holzenberger},
  \bibinfo{person}{Noam Kolt}, \bibinfo{person}{Peter Henderson},
  \bibinfo{person}{Sean Rehaag}, \bibinfo{person}{Sharad Goel},
  \bibinfo{person}{Shang Gao}, \bibinfo{person}{Spencer Williams},
  \bibinfo{person}{Sunny Gandhi}, \bibinfo{person}{Tom Zur},
  \bibinfo{person}{Varun Iyer}, {and} \bibinfo{person}{Zehua Li}.}
  \bibinfo{year}{2023}\natexlab{}.
\newblock \bibinfo{title}{LegalBench: A Collaboratively Built Benchmark for
  Measuring Legal Reasoning in Large Language Models}.
\newblock
\urldef\tempurl%
\url{https://arxiv.org/abs/2308.11462}
\showURL{%
\tempurl}


\bibitem[Hao et~al\mbox{.}(2026)]%
        {hao_2026_artificial}
\bibfield{author}{\bibinfo{person}{Qianyue Hao}, \bibinfo{person}{Fengli Xu},
  \bibinfo{person}{Yong Li}, {and} \bibinfo{person}{James Evans}.}
  \bibinfo{year}{2026}\natexlab{}.
\newblock \showarticletitle{Artificial intelligence tools expand scientists’
  impact but contract science’s focus}.
\newblock \bibinfo{journal}{\emph{Nature}} (\bibinfo{date}{01}
  \bibinfo{year}{2026}).
\newblock
\href{https://doi.org/10.1038/s41586-025-09922-y}{doi:\nolinkurl{10.1038/s41586-025-09922-y}}


\bibitem[Huang et~al\mbox{.}(2024)]%
        {huang_2024_a}
\bibfield{author}{\bibinfo{person}{Lei Huang}, \bibinfo{person}{Weijiang Yu},
  \bibinfo{person}{Weitao Ma}, \bibinfo{person}{Weihong Zhong},
  \bibinfo{person}{Zhangyin Feng}, \bibinfo{person}{Haotian Wang},
  \bibinfo{person}{Qianglong Chen}, \bibinfo{person}{Weihua Peng},
  \bibinfo{person}{Xiaocheng Feng}, \bibinfo{person}{Bing Qin}, {and}
  \bibinfo{person}{Ting Liu}.} \bibinfo{year}{2024}\natexlab{}.
\newblock \showarticletitle{A Survey on Hallucination in Large Language Models:
  Principles, Taxonomy, Challenges, and Open Questions}.
\newblock \bibinfo{journal}{\emph{ACM transactions on office information
  systems}}  \bibinfo{volume}{43} (\bibinfo{date}{11} \bibinfo{year}{2024}).
\newblock
\href{https://doi.org/10.1145/3703155}{doi:\nolinkurl{10.1145/3703155}}


\bibitem[Huang et~al\mbox{.}(2025)]%
        {huang_2025_evaluating}
\bibfield{author}{\bibinfo{person}{Yu~Chin Huang}, \bibinfo{person}{Yuhan Ji},
  {and} \bibinfo{person}{Song Gao}.} \bibinfo{year}{2025}\natexlab{}.
\newblock \showarticletitle{Evaluating Geospatial Reasoning Capabilities in
  Large Language Models: A Benchmark on Geometry Classification, Topological
  Relations and Direction Estimation}.
\newblock \bibinfo{journal}{\emph{Proceedings of the 4th ACM SIGSPATIAL
  International Workshop on Spatial Big Data and AI for Industrial
  Applications}} (\bibinfo{date}{11} \bibinfo{year}{2025}),
  \bibinfo{pages}{64--71}.
\newblock
\href{https://doi.org/10.1145/3764919.3770881}{doi:\nolinkurl{10.1145/3764919.3770881}}


\bibitem[Jin et~al\mbox{.}(2021)]%
        {jin_2021_what}
\bibfield{author}{\bibinfo{person}{Di Jin}, \bibinfo{person}{Eileen Pan},
  \bibinfo{person}{Nassim Oufattole}, \bibinfo{person}{Wei-Hung Weng},
  \bibinfo{person}{Hanyi Fang}, {and} \bibinfo{person}{Peter Szolovits}.}
  \bibinfo{year}{2021}\natexlab{}.
\newblock \showarticletitle{What Disease Does This Patient Have? A Large-Scale
  Open Domain Question Answering Dataset from Medical Exams}.
\newblock \bibinfo{journal}{\emph{Applied Sciences}}  \bibinfo{volume}{11}
  (\bibinfo{date}{07} \bibinfo{year}{2021}), \bibinfo{pages}{6421}.
\newblock
\href{https://doi.org/10.3390/app11146421}{doi:\nolinkurl{10.3390/app11146421}}


\bibitem[Juhász(2024)]%
        {juhsz_2024_assessing}
\bibfield{author}{\bibinfo{person}{Levente Juhász}.}
  \bibinfo{year}{2024}\natexlab{}.
\newblock \showarticletitle{Assessing publication trends in selected GIScience
  journals}.
\newblock \bibinfo{journal}{\emph{International Journal of Geographical
  Information Science}}  \bibinfo{volume}{38} (\bibinfo{date}{05}
  \bibinfo{year}{2024}), \bibinfo{pages}{1443--1467}.
\newblock
\href{https://doi.org/10.1080/13658816.2024.2347306}{doi:\nolinkurl{10.1080/13658816.2024.2347306}}


\bibitem[Kadavath et~al\mbox{.}(2022)]%
        {kadavath_2022_language}
\bibfield{author}{\bibinfo{person}{Saurav Kadavath}, \bibinfo{person}{Tom
  Conerly}, \bibinfo{person}{Amanda Askell}, \bibinfo{person}{Tom Henighan},
  \bibinfo{person}{Dawn Drain}, \bibinfo{person}{Ethan Perez},
  \bibinfo{person}{Nicholas Schiefer}, \bibinfo{person}{Zac Hatfield-Dodds},
  \bibinfo{person}{Nova DasSarma}, \bibinfo{person}{Eli Tran-Johnson},
  \bibinfo{person}{Scott Johnston}, \bibinfo{person}{Sheer El-Showk},
  \bibinfo{person}{Andy Jones}, \bibinfo{person}{Nelson Elhage},
  \bibinfo{person}{Tristan Hume}, \bibinfo{person}{Anna Chen},
  \bibinfo{person}{Yuntao Bai}, \bibinfo{person}{Sam Bowman},
  \bibinfo{person}{Stanislav Fort}, \bibinfo{person}{Deep Ganguli},
  \bibinfo{person}{Danny Hernandez}, \bibinfo{person}{Josh Jacobson},
  \bibinfo{person}{Jackson Kernion}, \bibinfo{person}{Shauna Kravec},
  \bibinfo{person}{Liane Lovitt}, \bibinfo{person}{Kamal Ndousse},
  \bibinfo{person}{Catherine Olsson}, \bibinfo{person}{Sam Ringer},
  \bibinfo{person}{Dario Amodei}, \bibinfo{person}{Tom Brown},
  \bibinfo{person}{Jack Clark}, \bibinfo{person}{Nicholas Joseph},
  \bibinfo{person}{Ben Mann}, \bibinfo{person}{Sam McCandlish},
  \bibinfo{person}{Chris Olah}, {and} \bibinfo{person}{Jared Kaplan}.}
  \bibinfo{year}{2022}\natexlab{}.
\newblock \bibinfo{title}{Language Models (Mostly) Know What They Know}.
\newblock
\urldef\tempurl%
\url{https://arxiv.org/abs/2207.05221}
\showURL{%
\tempurl}


\bibitem[Li et~al\mbox{.}(2024)]%
        {li_2024_streetviewllm}
\bibfield{author}{\bibinfo{person}{Zongrong Li}, \bibinfo{person}{Junhao Xu},
  \bibinfo{person}{Siqin Wang}, \bibinfo{person}{Yifan Wu}, {and}
  \bibinfo{person}{Haiyang Li}.} \bibinfo{year}{2024}\natexlab{}.
\newblock \bibinfo{title}{StreetviewLLM: Extracting Geographic Information
  Using a Chain-of-Thought Multimodal Large Language Model}.
\newblock
\urldef\tempurl%
\url{https://arxiv.org/abs/2411.14476}
\showURL{%
\tempurl}


\bibitem[Liang et~al\mbox{.}(2022)]%
        {liang_2022_holistic}
\bibfield{author}{\bibinfo{person}{Percy Liang}, \bibinfo{person}{Rishi
  Bommasani}, \bibinfo{person}{Tony Lee}, \bibinfo{person}{Dimitris Tsipras},
  \bibinfo{person}{Dilara Soylu}, \bibinfo{person}{Michihiro Yasunaga},
  \bibinfo{person}{Yian Zhang}, \bibinfo{person}{Deepak Narayanan},
  \bibinfo{person}{Yuhuai Wu}, \bibinfo{person}{Ananya Kumar},
  \bibinfo{person}{Benjamin Newman}, \bibinfo{person}{Binhang Yuan},
  \bibinfo{person}{Bobby Yan}, \bibinfo{person}{Ce Zhang},
  \bibinfo{person}{Christian Cosgrove}, \bibinfo{person}{Christopher~D.
  Manning}, \bibinfo{person}{Christopher Ré}, \bibinfo{person}{Diana
  Acosta-Navas}, \bibinfo{person}{Drew~A. Hudson}, \bibinfo{person}{Eric
  Zelikman}, \bibinfo{person}{Esin Durmus}, \bibinfo{person}{Faisal Ladhak},
  \bibinfo{person}{Frieda Rong}, \bibinfo{person}{Hongyu Ren},
  \bibinfo{person}{Huaxiu Yao}, \bibinfo{person}{Jue Wang},
  \bibinfo{person}{Keshav Santhanam}, \bibinfo{person}{Laurel Orr},
  \bibinfo{person}{Lucia Zheng}, \bibinfo{person}{Mert Yuksekgonul},
  \bibinfo{person}{Mirac Suzgun}, \bibinfo{person}{Nathan Kim},
  \bibinfo{person}{Neel Guha}, \bibinfo{person}{Niladri Chatterji},
  \bibinfo{person}{Omar Khattab}, \bibinfo{person}{Peter Henderson},
  \bibinfo{person}{Qian Huang}, \bibinfo{person}{Ryan Chi},
  \bibinfo{person}{Sang~Michael Xie}, \bibinfo{person}{Shibani Santurkar},
  \bibinfo{person}{Surya Ganguli}, \bibinfo{person}{Tatsunori Hashimoto},
  \bibinfo{person}{Thomas Icard}, \bibinfo{person}{Tianyi Zhang},
  \bibinfo{person}{Vishrav Chaudhary}, \bibinfo{person}{William Wang},
  \bibinfo{person}{Xuechen Li}, \bibinfo{person}{Yifan Mai},
  \bibinfo{person}{Yuhui Zhang}, {and} \bibinfo{person}{Yuta Koreeda}.}
  \bibinfo{year}{2022}\natexlab{}.
\newblock \bibinfo{title}{Holistic Evaluation of Language Models}.
\newblock
\urldef\tempurl%
\url{https://arxiv.org/abs/2211.09110}
\showURL{%
\tempurl}


\bibitem[Liang et~al\mbox{.}(2025)]%
        {liang_2025_quantifying}
\bibfield{author}{\bibinfo{person}{Weixin Liang}, \bibinfo{person}{Yaohui
  Zhang}, \bibinfo{person}{Zhengxuan Wu}, \bibinfo{person}{Haley Lepp},
  \bibinfo{person}{Wenlong Ji}, \bibinfo{person}{Xuandong Zhao},
  \bibinfo{person}{Hancheng Cao}, \bibinfo{person}{Sheng Liu},
  \bibinfo{person}{Siyu He}, \bibinfo{person}{Zhi Huang}, \bibinfo{person}{Diyi
  Yang}, \bibinfo{person}{Christopher Potts}, \bibinfo{person}{Christopher~D
  Manning}, {and} \bibinfo{person}{James Zou}.}
  \bibinfo{year}{2025}\natexlab{}.
\newblock \showarticletitle{Quantifying large language model usage in
  scientific papers}.
\newblock \bibinfo{journal}{\emph{Nature Human Behaviour}} (\bibinfo{date}{08}
  \bibinfo{year}{2025}).
\newblock
\href{https://doi.org/10.1038/s41562-025-02273-8}{doi:\nolinkurl{10.1038/s41562-025-02273-8}}


\bibitem[Lin et~al\mbox{.}(2022)]%
        {lin_2022_truthfulqa}
\bibfield{author}{\bibinfo{person}{Stephanie Lin}, \bibinfo{person}{Jacob
  Hilton}, {and} \bibinfo{person}{Owain Evans}.}
  \bibinfo{year}{2022}\natexlab{}.
\newblock \showarticletitle{TruthfulQA: Measuring How Models Mimic Human
  Falsehoods}.
\newblock \bibinfo{journal}{\emph{Proceedings of the 60th Annual Meeting of the
  Association for Computational Linguistics (Volume 1: Long Papers)}}
  (\bibinfo{date}{01} \bibinfo{year}{2022}).
\newblock
\href{https://doi.org/10.18653/v1/2022.acl-long.229}{doi:\nolinkurl{10.18653/v1/2022.acl-long.229}}


\bibitem[Luo et~al\mbox{.}(2025)]%
        {luo_2025_llm4sr}
\bibfield{author}{\bibinfo{person}{Ziming Luo}, \bibinfo{person}{Zonglin Yang},
  \bibinfo{person}{Zexin Xu}, \bibinfo{person}{Wei Yang}, {and}
  \bibinfo{person}{Xinya Du}.} \bibinfo{year}{2025}\natexlab{}.
\newblock \bibinfo{title}{LLM4SR: A Survey on Large Language Models for
  Scientific Research}.
\newblock
\urldef\tempurl%
\url{https://arxiv.org/abs/2501.04306}
\showURL{%
\tempurl}


\bibitem[Mai et~al\mbox{.}(2024)]%
        {mai_2024_on}
\bibfield{author}{\bibinfo{person}{Gengchen Mai}, \bibinfo{person}{Weiming
  Huang}, \bibinfo{person}{Jin Sun}, \bibinfo{person}{Suhang Song},
  \bibinfo{person}{Deepak Mishra}, \bibinfo{person}{Ninghao Liu},
  \bibinfo{person}{Song Gao}, \bibinfo{person}{Tianming Liu},
  \bibinfo{person}{Gao Cong}, \bibinfo{person}{Yingjie Hu},
  \bibinfo{person}{Chris Cundy}, \bibinfo{person}{Ziyuan Li},
  \bibinfo{person}{Rui Zhu}, {and} \bibinfo{person}{Ni Lao}.}
  \bibinfo{year}{2024}\natexlab{}.
\newblock \showarticletitle{On the Opportunities and Challenges of Foundation
  Models for GeoAI (Vision Paper)}.
\newblock \bibinfo{journal}{\emph{ACM transactions on spatial algorithms and
  systems}} (\bibinfo{date}{03} \bibinfo{year}{2024}).
\newblock
\href{https://doi.org/10.1145/3653070}{doi:\nolinkurl{10.1145/3653070}}


\bibitem[Manvi et~al\mbox{.}(2023)]%
        {manvi_2023_geollm}
\bibfield{author}{\bibinfo{person}{Rohin Manvi}, \bibinfo{person}{Samar
  Khanna}, \bibinfo{person}{Gengchen Mai}, \bibinfo{person}{Marshall Burke},
  \bibinfo{person}{David Lobell}, {and} \bibinfo{person}{Stefano Ermon}.}
  \bibinfo{year}{2023}\natexlab{}.
\newblock \bibinfo{title}{GeoLLM: Extracting Geospatial Knowledge from Large
  Language Models}.
\newblock
\urldef\tempurl%
\url{https://arxiv.org/abs/2310.06213}
\showURL{%
\tempurl}


\bibitem[Min et~al\mbox{.}(2023)]%
        {min_2023_factscore}
\bibfield{author}{\bibinfo{person}{Sewon Min}, \bibinfo{person}{Kalpesh
  Krishna}, \bibinfo{person}{Xinxi Lyu}, \bibinfo{person}{Mike Lewis},
  \bibinfo{person}{Wen-tau Yih}, \bibinfo{person}{Pang Koh},
  \bibinfo{person}{Mohit Iyyer}, \bibinfo{person}{Luke Zettlemoyer}, {and}
  \bibinfo{person}{Hannaneh Hajishirzi}.} \bibinfo{year}{2023}\natexlab{}.
\newblock \showarticletitle{FActScore: Fine-grained Atomic Evaluation of
  Factual Precision in Long Form Text Generation}.
\newblock \bibinfo{journal}{\emph{In Proceedings of the 2023 Conference on
  Empirical Methods in Natural Language Processing (EMNLP 2023)}}
  (\bibinfo{date}{01} \bibinfo{year}{2023}).
\newblock
\href{https://doi.org/10.18653/v1/2023.emnlp-main.741}{doi:\nolinkurl{10.18653/v1/2023.emnlp-main.741}}


\bibitem[Mugaanyi et~al\mbox{.}(2023)]%
        {mugaanyi_2023_citations}
\bibfield{author}{\bibinfo{person}{Joseph Mugaanyi}, \bibinfo{person}{Liuying
  Cai}, \bibinfo{person}{Sumei Cheng}, \bibinfo{person}{Caide Lu}, {and}
  \bibinfo{person}{Jing Huang}.} \bibinfo{year}{2023}\natexlab{}.
\newblock \showarticletitle{Citations and References in Scholarly Writing: A
  cross-disciplinary Evaluation of Large Language Model Performance and
  Reliability. (Preprint)}.
\newblock \bibinfo{journal}{\emph{JMIR. Journal of medical internet
  research/Journal of medical internet research}}  \bibinfo{volume}{26}
  (\bibinfo{date}{09} \bibinfo{year}{2023}).
\newblock
\href{https://doi.org/10.2196/52935}{doi:\nolinkurl{10.2196/52935}}


\bibitem[Scherbakov et~al\mbox{.}(2025)]%
        {scherbakov_2025_the}
\bibfield{author}{\bibinfo{person}{Dmitry Scherbakov}, \bibinfo{person}{Nina
  Hubig}, \bibinfo{person}{Vinita Jansari}, \bibinfo{person}{Alexander
  Bakumenko}, {and} \bibinfo{person}{Leslie~A. Lenert}.}
  \bibinfo{year}{2025}\natexlab{}.
\newblock \showarticletitle{The emergence of large language models as tools in
  literature reviews: a large language model-assisted systematic review}.
\newblock \bibinfo{journal}{\emph{Journal of the American Medical Informatics
  Association}} \bibinfo{volume}{32}, \bibinfo{number}{6} (\bibinfo{date}{03}
  \bibinfo{year}{2025}), \bibinfo{pages}{1071--1086}.
\newblock
\href{https://doi.org/10.1093/jamia/ocaf063}{doi:\nolinkurl{10.1093/jamia/ocaf063}}


\bibitem[Sun et~al\mbox{.}(2024)]%
        {sun_2024_scieval}
\bibfield{author}{\bibinfo{person}{Liangtai Sun}, \bibinfo{person}{Yang Han},
  \bibinfo{person}{Zihan Zhao}, \bibinfo{person}{Da Ma},
  \bibinfo{person}{Zhennan Shen}, \bibinfo{person}{Baocai Chen},
  \bibinfo{person}{Lu Chen}, {and} \bibinfo{person}{Kai Yu}.}
  \bibinfo{year}{2024}\natexlab{}.
\newblock \showarticletitle{SciEval: A Multi-Level Large Language Model
  Evaluation Benchmark for Scientific Research}.
\newblock \bibinfo{journal}{\emph{Proceedings of the AAAI Conference on
  Artificial Intelligence}}  \bibinfo{volume}{38} (\bibinfo{date}{03}
  \bibinfo{year}{2024}), \bibinfo{pages}{19053--19061}.
\newblock
\href{https://doi.org/10.1609/aaai.v38i17.29872}{doi:\nolinkurl{10.1609/aaai.v38i17.29872}}


\bibitem[Topaz et~al\mbox{.}(2026)]%
        {topaz_2026_fabricated}
\bibfield{author}{\bibinfo{person}{Maxim Topaz}, \bibinfo{person}{Nir Roguin},
  \bibinfo{person}{Pallavi Gupta}, \bibinfo{person}{Zhihong Zhang}, {and}
  \bibinfo{person}{Laura-Maria Peltonen}.} \bibinfo{year}{2026}\natexlab{}.
\newblock \showarticletitle{Fabricated citations: an audit across 2·5 million
  biomedical papers}.
\newblock \bibinfo{journal}{\emph{The Lancet}}  \bibinfo{volume}{407}
  (\bibinfo{date}{05} \bibinfo{year}{2026}), \bibinfo{pages}{1779--1781}.
\newblock
\href{https://doi.org/10.1016/s0140-6736(26)00603-3}{doi:\nolinkurl{10.1016/s0140-6736(26)00603-3}}


\bibitem[Walters and Wilder(2023)]%
        {walters_2023_fabrication}
\bibfield{author}{\bibinfo{person}{William~H. Walters} {and}
  \bibinfo{person}{Esther~Isabelle Wilder}.} \bibinfo{year}{2023}\natexlab{}.
\newblock \showarticletitle{Fabrication and errors in the bibliographic
  citations generated by ChatGPT}.
\newblock \bibinfo{journal}{\emph{Scientific Reports}}  \bibinfo{volume}{13}
  (\bibinfo{date}{09} \bibinfo{year}{2023}), \bibinfo{pages}{14045}.
\newblock
\href{https://doi.org/10.1038/s41598-023-41032-5}{doi:\nolinkurl{10.1038/s41598-023-41032-5}}


\bibitem[Xiong et~al\mbox{.}(2023)]%
        {xiong_2023_can}
\bibfield{author}{\bibinfo{person}{Miao Xiong}, \bibinfo{person}{Zhiyuan Hu},
  \bibinfo{person}{Xinyang Lu}, \bibinfo{person}{Yifei Li},
  \bibinfo{person}{Jie Fu}, \bibinfo{person}{Junxian He}, {and}
  \bibinfo{person}{Bryan Hooi}.} \bibinfo{year}{2023}\natexlab{}.
\newblock \bibinfo{title}{Can LLMs Express Their Uncertainty? An Empirical
  Evaluation of Confidence Elicitation in LLMs}.
\newblock
\urldef\tempurl%
\url{https://arxiv.org/abs/2306.13063}
\showURL{%
\tempurl}


\end{thebibliography}

\appendix

\section{Prompt Templates}

\begin{figure*}[htbp]
  \centering
  \includegraphics[width=\textwidth,height=0.88\textheight,keepaspectratio]{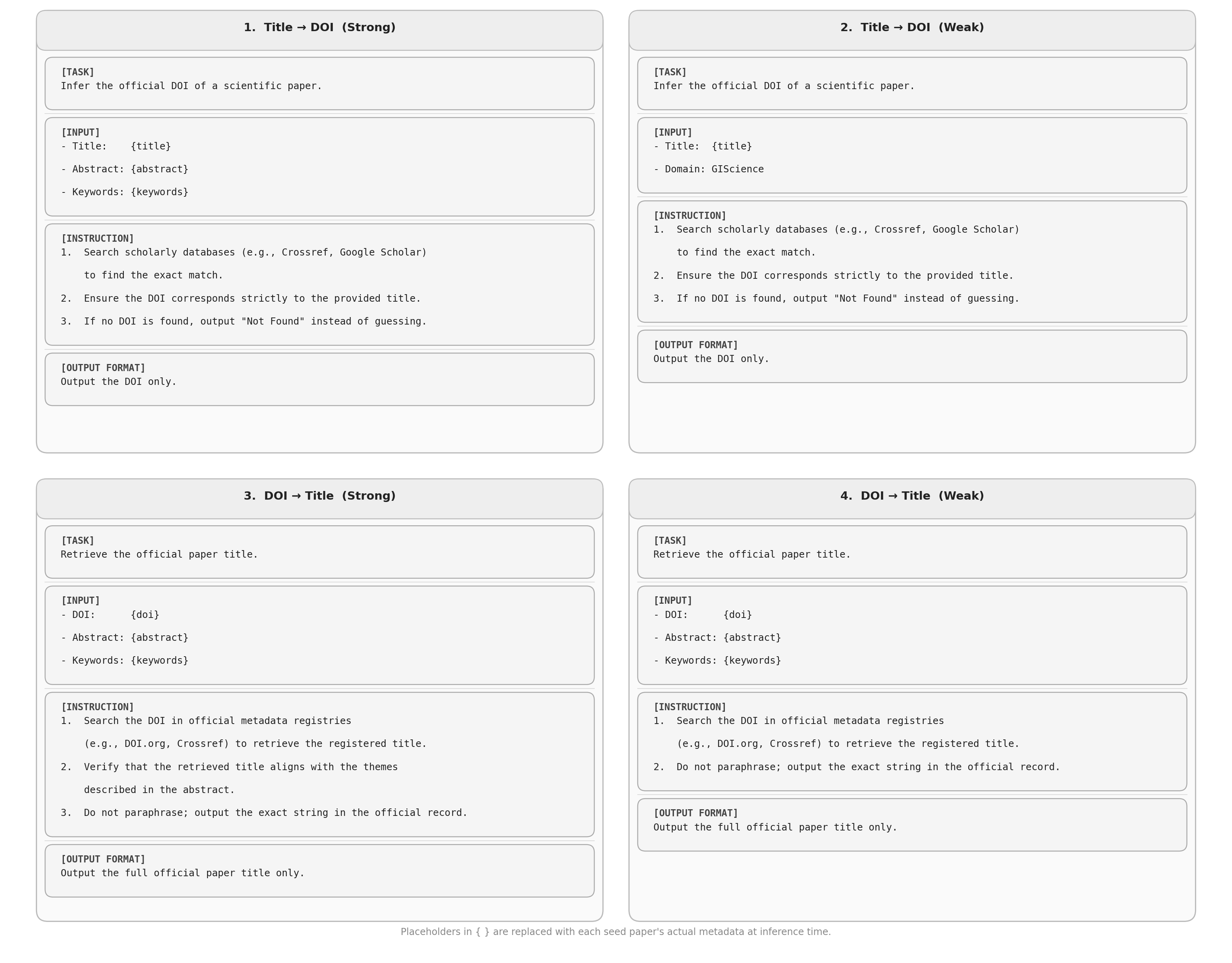}
  \caption{Prompt Templates for Task~1 Metadata Retrieval Evaluation.}
  \label{fig:promptA1}
\end{figure*}

\begin{figure*}[htbp]
  \centering
  \includegraphics[width=\textwidth,height=0.54\textheight,keepaspectratio]{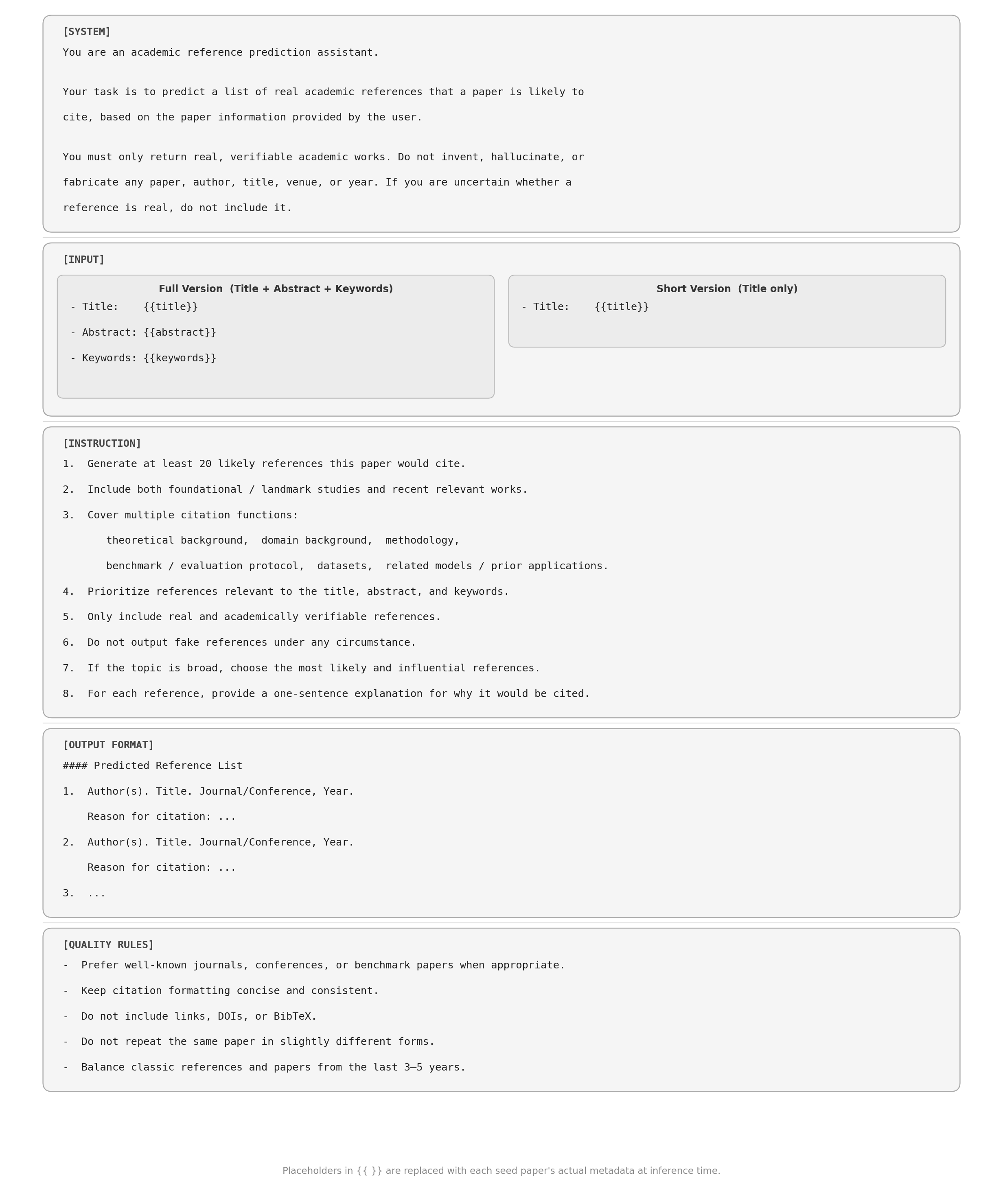}\\[6pt]
  \includegraphics[width=\textwidth,height=0.34\textheight,keepaspectratio]{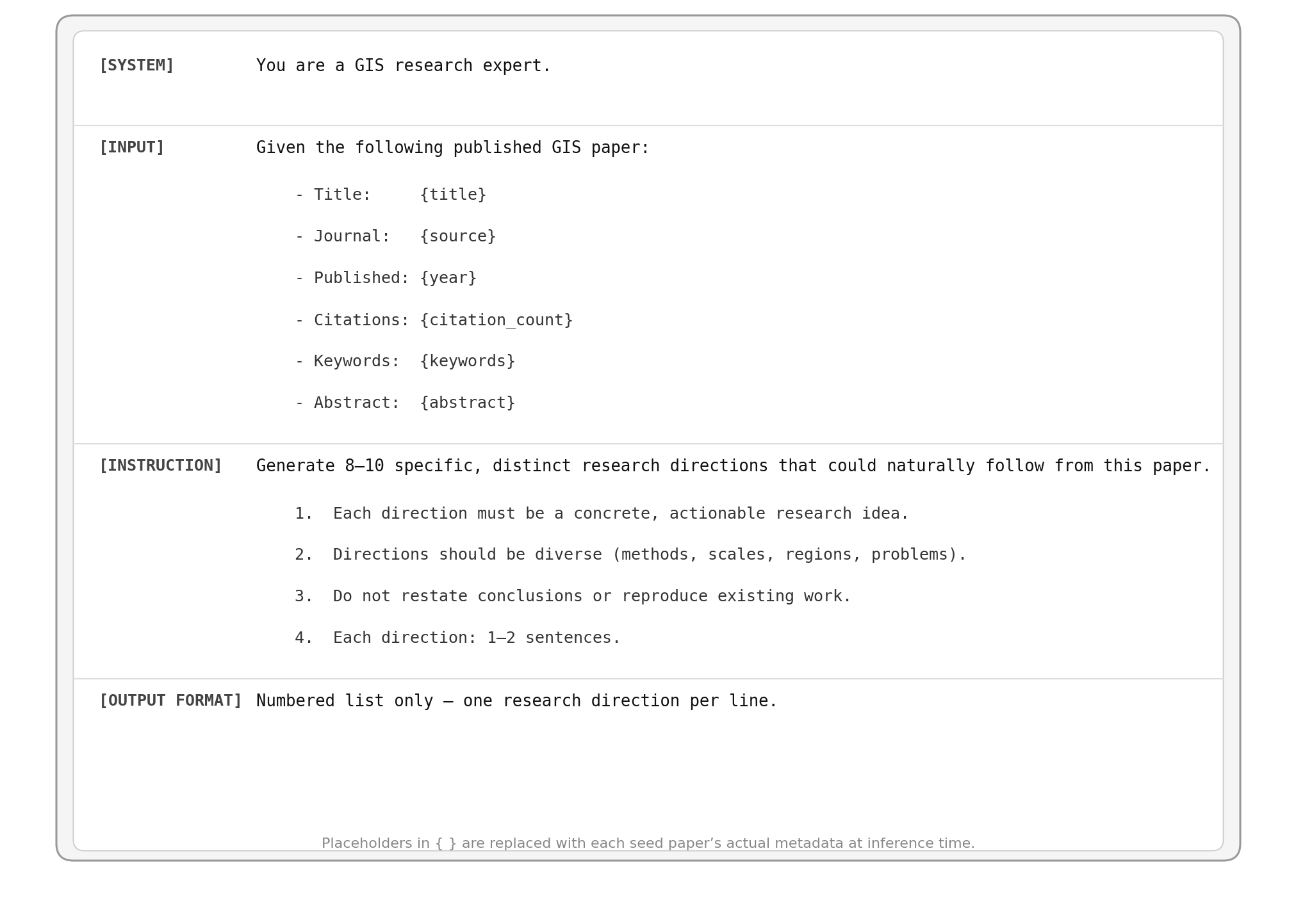}
  \caption{Prompt Templates for Task~2 Literature Linking (top) and Task~3 Research Direction Generation (bottom).}
  \label{fig:promptA2}
\end{figure*}

\end{document}